\documentclass[journal]{IEEEtran}
\usepackage{siunitx}
\usepackage{amsmath}
\usepackage[switch]{lineno}

\usepackage{hyperref} 

\ifCLASSINFOpdf
    \usepackage[pdftex]{graphicx}
    \graphicspath{{./fig/}}
    \DeclareGraphicsExtensions{.pdf,.jpeg,.png}
\else
\fi

\usepackage{xargs}
\usepackage{ifdraft}
\usepackage{changepage}
\usepackage[colorinlistoftodos,prependcaption,textsize=large]{todonotes}

\newcommandx{\marginnote}[2][1=]{\todo[linecolor=blue,backgroundcolor=blue!25,bordercolor=blue,#1]{#2}}
\newcommandx{\thiswillnotshow}[2][1=]{\marginnote[disable,#1]{#2}}

\begin{document}
\bstctlcite{IEEEexample:BSTcontrol}
\title{Simulation Study of Photon-to-Digital Converter (PDC) Timing Specifications for LoLX Experiment}
\author{
    Nguyen V. H. Viet,~\IEEEmembership{Member,~IEEE,}
    Alaa Al Masri,
    Masaharu Nomachi,~\IEEEmembership{Senior~Member,~IEEE,}
    Marc-André~Tétrault,~\IEEEmembership{Member,~IEEE,}
    Soud Al Kharusi,
    Thomas Brunner,
    Christopher Chambers,
    Bindiya Chana,
    Austin de St. Croix,
    Eamon Egan,
    Marco Francesconi,
    David Gallacher,
    Luca Galli,
    Pietro Giampa,
    Damian Goeldi,
    Jessee Lefebvre,
    Chloe Malbrunot,
    Peter Margetak,
    Juliette Martin,
    Thomas McElroy,
    Mayur Patel,
    Bernadette Rebeiro,
    Fabrice~Retière,
    El Mehdi Rtimi,
    Lisa Rudolph,
    Simon Viel,
    Liang Xie
    \thanks{Manuscript received October 20, 2023;
    revised August 30, 2024; 
    This research was undertaken thanks in part to funding from the Canada First Research Excellence Fund through the Arthur B. McDonald Astroparticle Physics Research Institute, with support from the National Sciences and Engineering Research Council of Canada (NSERC), and the Fonds de Recherche du Quebec Nature et Technologies (FQRNT). This was also supported by CHELN and CHELNX projects funded by Grant INFN n. 19593, and the PQBA program of Osaka University.
    
    N. V. H. Viet and M. Nomachi are with the Research Center for Nuclear Physics (RCNP), Osaka University, Osaka, Japan. (e-mail: nvhviet@rcnp.osaka-u.ac.jp)
    
    A. Al Masri, M.-A.Tétrault, J. Lefebvre, and E. M. Rtimi are with the Department of Electrical and Computer Engineering, Université de Sherbrooke, Sherbrooke, QC, Canada.
    
    S. Al Kharusi, T. Brunner, C. Chambers, E. Egan, D. Gallacher, T. McElroy, B. Rebeiro, and L. Rudolph are with the Physics Department, McGill University, Montréal, QC, Canada.
    
    B. Chana, D. Goeldi, and S. Viel are with the Department of Physics, Carleton University, Ottawa, ON, Canada.
    
    A. de St. Croix, C. Malbrunot, P. Margetak, J. Martin, M. Patel, F. Retière, and L. Xie are with TRIUMF, Vancouver, BC, Canada.
    
    M. Francesconi and L. Galli are with INFN, Pisa, Italy.
    
    P. Giampa is with SNOLAB, Lively, ON, Canada.
    
    }
}
\IEEEaftertitletext{\vspace{-1.5\baselineskip}}
\maketitle
\begin{abstract}
The Light only Liquid Xenon (LoLX) experiment is a prototype detector aimed to study liquid xenon (LXe) light properties and various photodetection technologies.
LoLX is also aimed to quantify LXe's time resolution as a potential scintillator for 10~ps time-of-flight positron emission tomography (TOF-PET).
Another key goal of LoLX is to perform a time-based separation of Cerenkov and scintillation photons for new background rejection methods in LXe experiments.
To achieve this separation, LoLX is set to be equipped with photon-to-digital converters (PDCs), a photosensor type that can provide a timestamp for each observed photon. 
To guide the PDC design, we explore the requirements and potential outcomes for time-based Cerenkov separation.
We use a PDC simulator, whose input is the light information from the Geant4-based LoLX simulation model, and evaluate the separation quality against time-to-digital converter (TDC) parameters of the PDCs.

Compared with the current filter-based approach, the simulations predict a few different configurations that offer a Cerenkov separation level increase from 50\% to 66\% when using PDCs and time-based separation.
A separation of 65\% is also achievable with just 16 TDCs for 14,400 micro-cells per PDC, or one TDC per 2.25~mm$^2$.
These simulation results will lead to a specification guide for the upcoming PDC design as well as expected results to compare against future PDC-based experimental measurements. 
In the longer term, the overall LoLX results will assist large LXe-based experiments and motivate the assembly of a LXe-based TOF-PET demonstrator system.
\end{abstract}

\section{Introduction} \label{intro}
\IEEEPARstart{S}{cintillators} are core components for many particle physics detectors.
The choice of scintillator relies on its material properties and the sought-after experimental data, generally position, deposited energy and occurrence time.
Liquid xenon (LXe) is a scintillator offering attractive performances on these three figures of merit due to its good light yield, fast timing, and dual light/charge readout paths. 
Although it needs to be cooled to -110~\textcelsius, and thus requires a cryostat to operate, it can be shaped into a large, continuous volume, an attractive feature for neutrino experiments like nEXO \cite{adhikari2021nexo}, which requires 1\% energy resolution at 2.5~MeV Q-value of $^{136}$Xe in its search for neutrinoless double beta decay ($0\nu2\beta$). 
The Light-only Liquid Xenon (LoLX) experiment \cite{galli2023looking, de2020light} supports nEXO in testing candidate photosensor technologies and studying LXe light properties to achieve this goal.
In a future phase of the experiment, LoLX will aim to achieve time-based Cerenkov-scintillation separation to explore new background rejection methods in LXe-based $0\nu2\beta$ experiments \cite{BRODSKY201976}.
Moreover, Cerenkov photons in LXe can potentially increase prompt photon statistics for fast timing applications such as time-of-flight positron emission tomography (TOF-PET) \cite{Gundacker_2016, PETALO2020, doke-2006-TOF-PET-LXe}.
For this application, LoLX’s goal is to confirm if a 10~ps time resolution can be obtained in LXe to pave the way for PET scanners with cutting edge performance \cite{Lecoq_2020}.

To achieve those goals, current silicon photo-multipliers (SiPMs) installed in LoLX are expected to be replaced by photon-to-digital converters (PDCs) \cite{Pratte_PDC_2021}, also known as digital SiPMs.
Similar to SiPM, a PDC consists of thousands of single photon avalanche diodes (SPADs), called pixels or micro-cells.
However, instead of converting the SPADs' Boolean information to into an analog current sum as done in SiPMs, PDCs have each SPAD or group of SPADs directly timestamped by a time-to-digital converter (TDC). 
Through this approach, PDCs promise to offer sub-100~ps timing for each observed photon and to address the intrinsic timing skew limitation of large area SiPMs \cite{Nolet_2016}.

In this study, the light output information from the LoLX simulation model \cite{chana2023PhDthesis, 2024-LoLX-ExCT} was redirected to a PDC simulator, the digital SPAD array simulator (DSAS) \cite{ACTherrien_DSAS_2014}. 
The work presented here, through the redirection of the optical photons to the DSAS model, explores the feasibility of time-based Cerenkov-scintillation separation on an event-by-event basis. This exploration then defines target PDC specifications and available margins to guide the design for the actual PDCs. 

\section{Setups \& Methods} \label{setups-methods}
\subsection{LoLX Detector} \label{lolx-det}
\begin{figure}[!t]
\centering
\includegraphics[width=\linewidth]{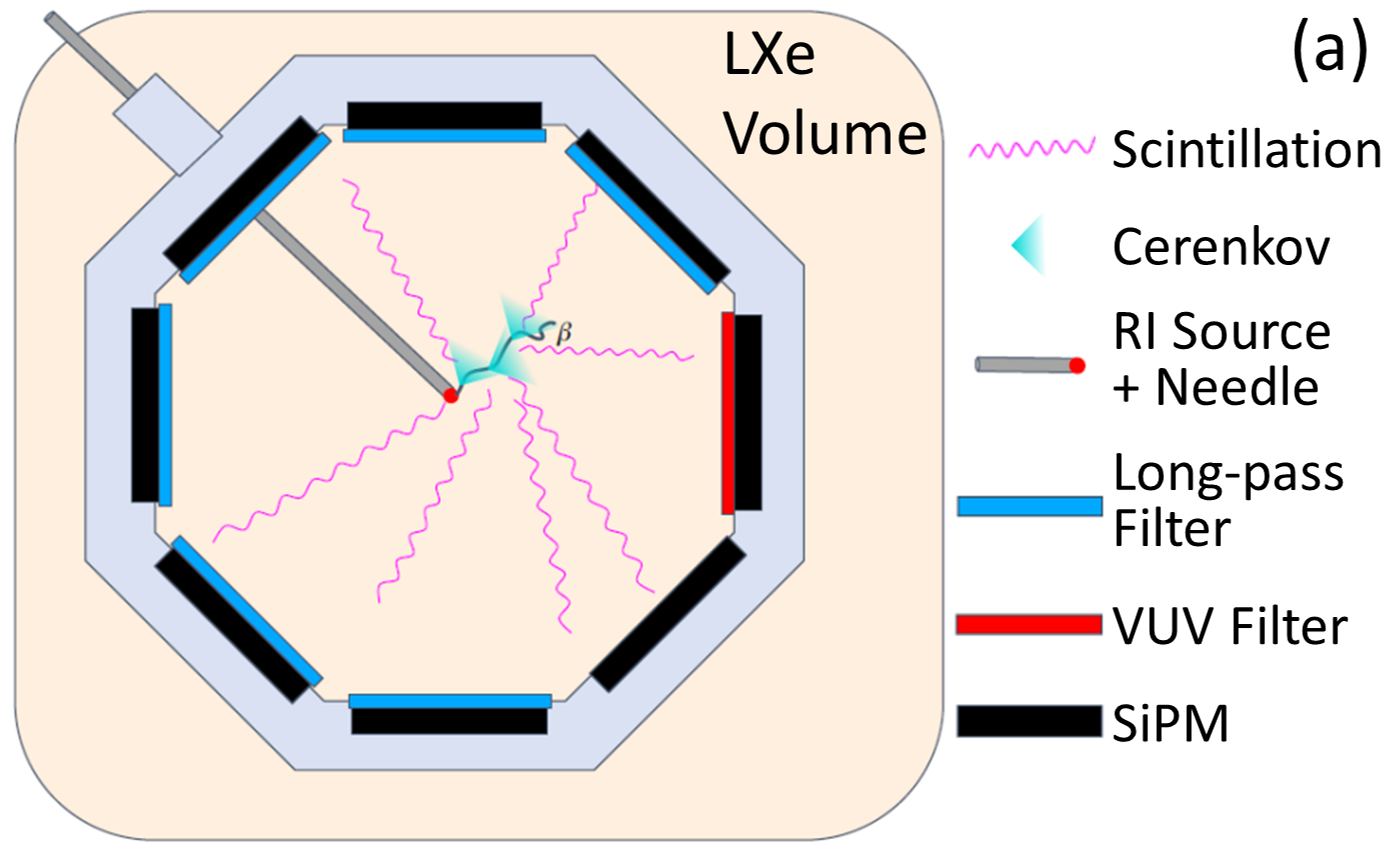}
\includegraphics[width=\linewidth]{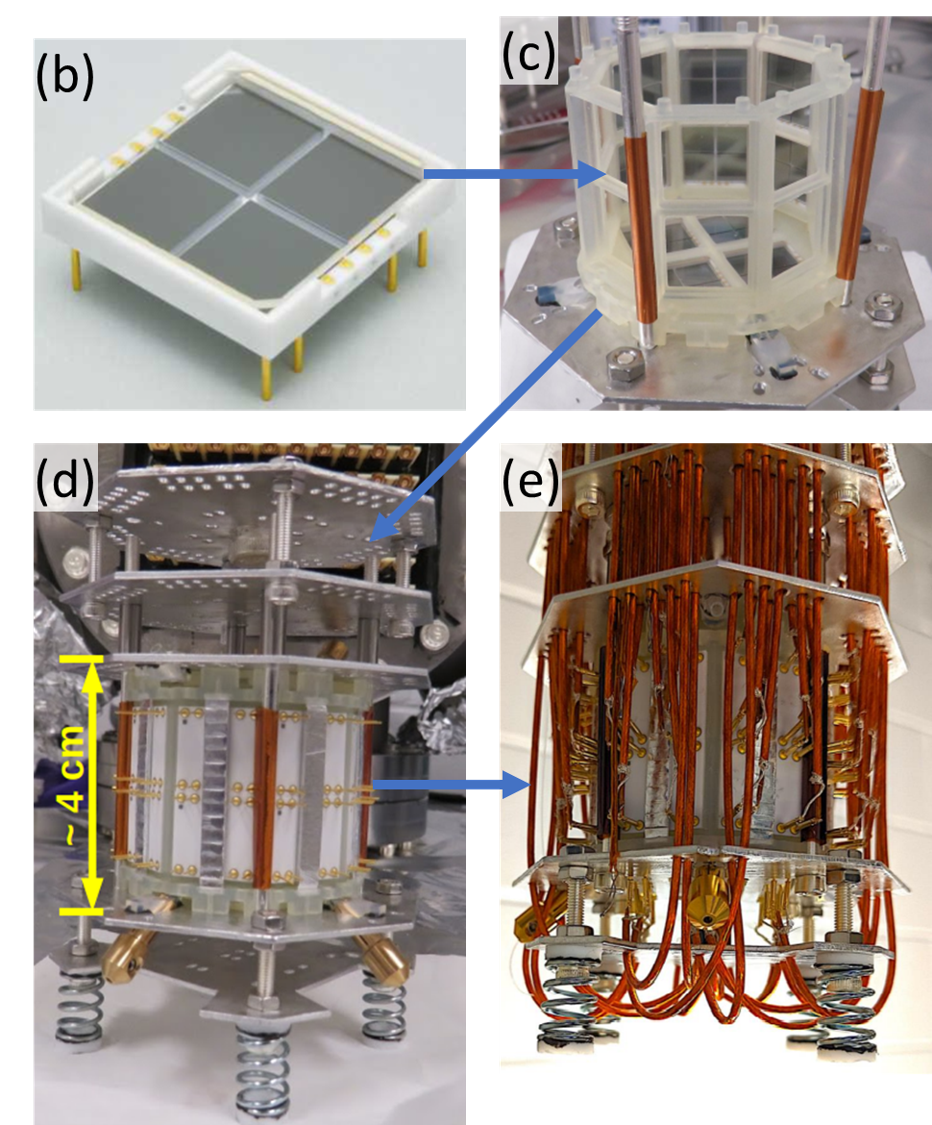}
\caption{
(a) Sketch of LoLX setup; 
(b) \qtyproduct{1.5x1.5}{\cm} VUV4 Quad SiPM module; 
(c) 3D printed cage for 24 modules; 
(d) assembled LoLX detector (unwired); 
(e) assembled LoLX detector (wired).
}
\label{fig_lolx}
\end{figure}

The LoLX detector cage is a 3D printed octagonal prism, enclosing a LXe volume of approximately 30~\unit{\cubic\cm} (Fig. \ref{fig_lolx}).
This cage is equipped with 24 Hamamatsu VUV4 Quad SiPM modules (part number: S13371-6050CQ) \cite{HPK-VUV4-MPPC-datasheet}, 4 SiPMs each, so 96 SiPMs in total.
The performance of this SiPM family was reported previously in \cite{2022-nEXO-VUV-SiPMs} and \cite{2019-nEXO-HPK-VUV4-MPPCs}.
To separate Cerenkov and scintillation, LoLX currently uses optical filters installed on the surface of the SiPM modules: 22 modules with long-pass filters (allow wavelength $>$~225~nm, mainly Cerenkov), 1 module with VUV filter (allow wavelength 150--180~nm, mainly scintillation), and 1 module with no filter (bare).
A Sr-90 source is inserted into the instrumented volume through a needle that penetrates the cage. 
The LoLX detector is enclosed in a cryostat located at McGill University, containing approximately 5~kg of LXe.
The cryostat can be cooled down to reach a temperature of 165~K at approximately 1~atm. 
Signals are digitized using CAEN V1740 module with a sampling rate of 62.5~MS/s. 
More details about the LoLX detector can be found in \cite{chana2023PhDthesis} and \cite{2024-LoLX-ExCT}.

\subsection{LoLX Simulation} \label{lolx-sim}
The LoLX simulation model \cite{chana2023PhDthesis, 2024-LoLX-ExCT} was developed using Geant4 toolkit \cite{Geant4_2016}. 
The NEST model \cite{nest2011} is used for LXe scintillation generation. 
Cerenkov generation is handled by G4Cerenkov process of Geant4 where the wavelength-dependent refractive index of LXe, main factor for Cerenkov yield, is obtained from literature \cite{LXe_RI}.

In addition to Cerenkov and scintillation photons from LXe, external cross-talk (ExCT) photons are also simulated in the LoLX Geant4 model. 
They are near-infrared photons produced in the avalanche process of a SPAD cell that escape the SiPM, travel through the scintillation medium, and are detected by a distant SiPM.
Additional physics classes and custom optical boundary classes were created in Geant4 to produce these ExCT photons.
To correctly reflect LoLX's experimental results, photon detection/reflection by SiPMs in LoLX simulation incorporated ExCT generation, photo-detection efficiency (PDE) provided by Hamamatsu \cite{HPK-VUV4-MPPC-datasheet}, and the measured SiPM's reflectivity \cite{nEXO-2020-SiPM-reflectance}. 
Details on the LoLX simulation environment can be found in \cite{chana2023PhDthesis}, and the of ExCT from SiPMs in LoLX can be found in \cite{2024-LoLX-ExCT}.


\subsection{PDC Simulator (DSAS) Setup} \label{dsas}
In this study, information of photons detected by the SiPM surfaces in LoLX Geant4 simulation model was recorded and passed on the PDC simulator, DSAS \cite{ACTherrien_DSAS_2014}.
The PDC geometry specifications in DSAS were based on S13371-6050CQ SiPMs: approximately \qtyproduct{6x6}{\mm} active area, and 50~\unit{\um} pitch.
Other important parameters that need to be set in DSAS are related to PDE, after-pulsing (AP), optical cross-talk (CT), and dark count (DC). 
As we used the photon information after detection by the SiPM surfaces in the Geant4 simulation model, PDE in DSAS was set to 100\% to avoid double-applying the PDE.
Performance of LoLX's SiPMs were previously evaluated, indicating low noise at the detector operating temperature (165~K): DC is approximately 2 counts for an event window of 3~\unit{\us}, AP and CT total probability is 8--15\% \cite{2022-nEXO-VUV-SiPMs}. 
In the early time window, the first few nanoseconds, the scintillation and Cerenkov photon densities are significantly higher than the above-mentioned noise.
Thus, AP, CT, and DC were not simulated in DSAS for this PDC timing specification study to simplify analysis.
With 100\% PDE and without noise (no AP, CT, and DC), DSAS retains 98.3\% of the total photons detected by the SiPM surfaces in the Geant4 model.
The 1.7\% photon loss is due to the PDC's physical modeling in DSAS, e.g. oblique-angle photons near the PDC's edge escape its finite volume or two photons hit the same SPAD within its dead-time. 
DSAS source code is also available for community use in \cite{DSAS-source-code}.


\subsection{Filter performance in LoLX Geant4 simulation model} \label{filter-performance}
We evaluated the performance of long-pass filters in LoLX Geant4 model to set a baseline separation performance to compare with the time-based approach using PDCs.
The   combinationin the Geant4 model were modified from the original setup, which follows the actual configuration mentioned in Section \ref{lolx-det}, to two different cases: all bare and all long-pass filters for 96 photosensors.
From these two cases, we could evaluate the number of Cerenkov (or scintillation) photons going (or leaking) through the filters and being detected.
In this filter-based approach, on average, 50\% of the total Cerenkov photons go through the long-pass filters and are detected.
The ratio of the number of scintillation photons ($n_\text{Scint}$) leaking through these filters to the number of Cerenkov photons ($n_\text{Ceren}$) going through is approximately 10\%.
Based on these values, separation conditions for the time-based approach will be defined in the next section.

\subsection{PDC Timing Specification Study} \label{PDC-timing-study}
The goal of this study is to investigate if it is possible to improve the Cerenkov separation using time-only measurements, without filters, and how the TDC parameters affect the separation quality.

Three TDC parameters are considered in this study: (a) jitter, the uncertainty in a TDC's timing measurement, (b) least significant bit (LSB), the smallest timing step a TDC can measure, and (c) the SPAD:TDC sharing ratio, how many SPADs share a TDC. 
This third parameter comes from the possibility to have one TDC per SPAD or for a small group of SPADs \cite{Pratte_PDC_2021}, a feature not possible in analog SiPMs. 
Various combinations of the three parameters were simulated, and for each setting, a time-based upper cut was applied to select an early time window that maximizes the $n_\text{Ceren}$ obtained while minimizes the $n_\text{Scint}$ remaining in this window.
The separation quality was first evaluated with both TDC jitter (FWHM) and LSB varied between 1 and 50~ps.
The initial SPAD:TDC ratio was set to 1:1, and was subsequently altered after finding optimal jitter and LSB values. 
These parameters are varied within the DSAS model, where its other photodetection parameters have already been explained in Section \ref{dsas}. 

For the time-based separation purpose, photon information input to DSAS was taken from LoLX Geant4 model with no filter applied, i.e. all bare photosensors.
The Sr-90 source was positioned at the center of the LoLX detector cage. 

\begin{figure}[!t]
\centering
\includegraphics[width=\linewidth]{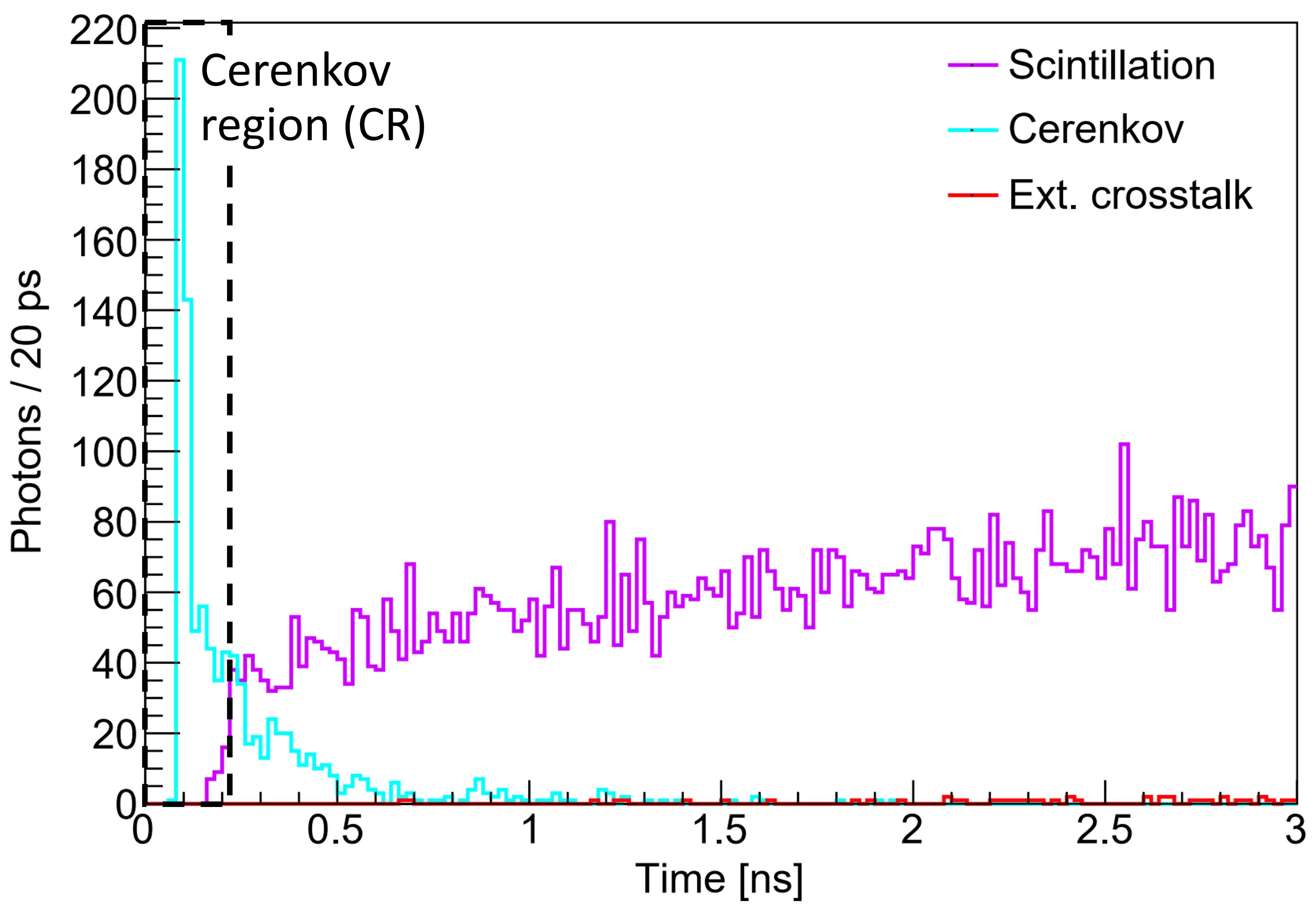}
\caption{Raw photon time profile of a bare photosensor from LoLX Geant4 simulation with focus on the early time window.
In LoLX simulation, from 5000 primary $\beta$ events, approximately 900 Cerenkov and 230,000 scintillation photons reach this bare photosensor.
After passing this photon profile into DSAS, the Cerenkov region as demonstrated in this figure should be selected to maximize the $n_\text{Ceren}$ and minimize the $n_\text{Scint}$ within it.
The ExCT photons as shown in this figure were treated as scintillation photons in DSAS due to their late arrival.
}
\label{fig_Cerenkov_Scint}
\end{figure}

Time-based cut conditions for the Cerenkov fraction and Scintillation-to-Cerenkov ratio are set to exceed the filter performance (obtained in Section \ref{filter-performance}), respectively, as follows:
\begin{equation}
    f_\text{Ceren} \ge 55\% \quad \text{\&} \quad r_\text{SC} < 5\% \tag{$*$}
\label{eq_cond}
\end{equation}
in which:
\begin{align*}
f_\text{Ceren} = \frac{n_\text{Ceren} \text{ in CR}}{\text{Total } n_\text{Ceren}} \quad \text{\&} \quad
r_\text{SC} = \frac{n_\text{Scint} \text{ remains in CR}}{n_\text{Ceren} \text{ in CR}}
\end{align*}
where CR is the Cerenkov Region as demonstrated in Fig.~\ref{fig_Cerenkov_Scint}.
While the $n_\text{Ceren}$ in this figure is much less than the $n_\text{Scint}$, it is possible to separate Cerenkov photons due to their prompt arrival.
For each photosensor, there could be several timing cut values meeting conditions (\ref{eq_cond}). 
The optimal cut value is defined as the one that maximizes the Cerenkov fraction, $f_\text{Ceren}$.
This optimal cut value is also the one with the latest time in all possible cut values.

Also present in the early time window of the raw photon time profile in Fig.~\ref{fig_Cerenkov_Scint} are a few ExCT photons mentioned in Section \ref{lolx-sim}.
Using the ExCT simulation model \cite{2024-LoLX-ExCT} for a no-filter setup of LoLX for this PDC study, ExCT photons account for around 7\% the total $n_\text{Scint}$.
As the presence of these ExCT photons in the early time window are rare, they nearly do not affect the Cerenkov-scintillation time-based separation.
Thus, ExCT photons were treated as scintillation photons in DSAS.


\section{Results \& Discussions} \label{result_discuss}
\subsection{TDC Jitter \& LSB} \label{jitter_LSB}
\begin{figure}[!t]
\centering
\includegraphics[width=\linewidth]{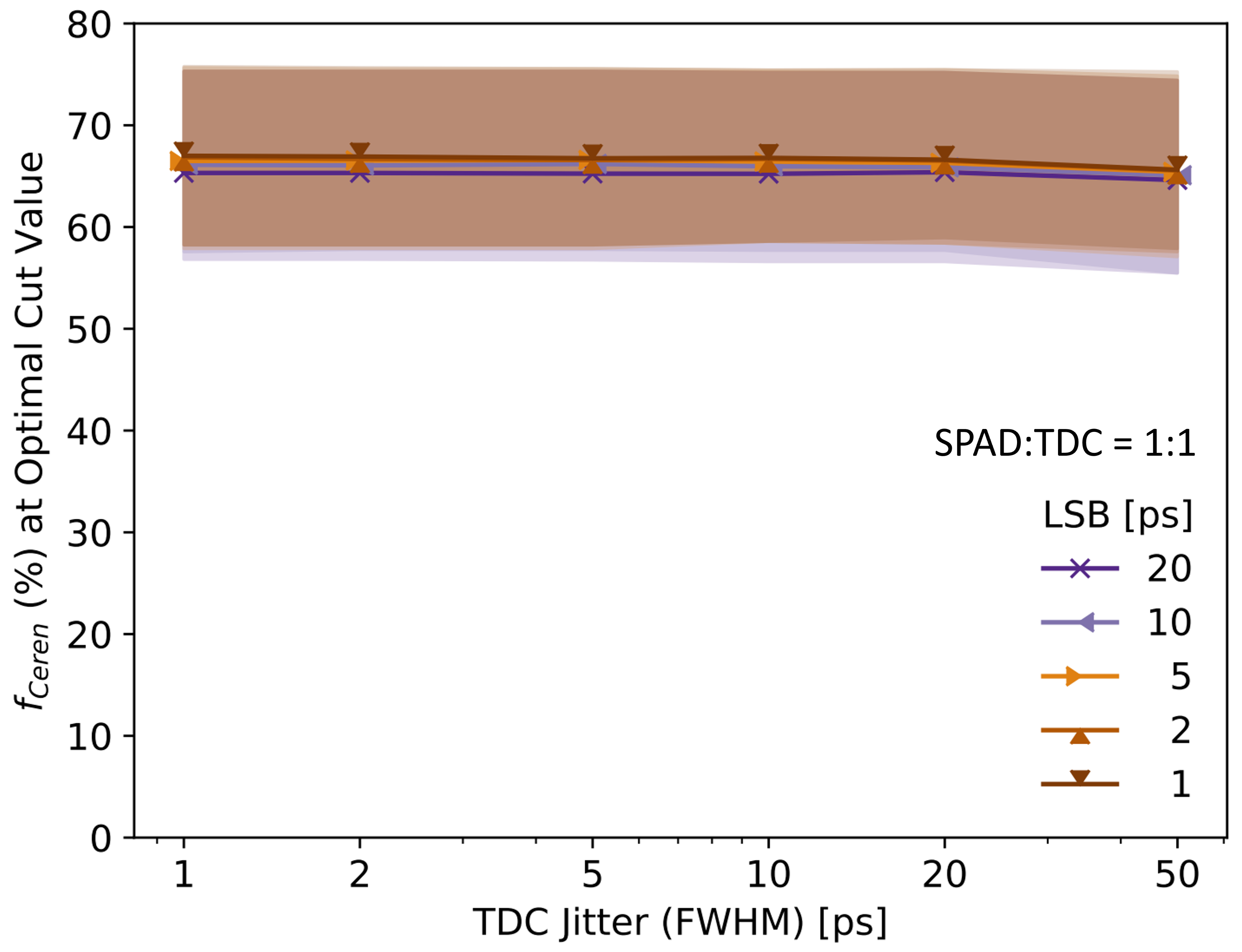}
\caption{
Cerenkov separation (as $f_\text{Ceren}$) of 96 photosensors vs. different TDC specifications with the optimal cut value for each photosensor.
Dots and lines depict medians while shaded areas show min and max values of 96 photosensors.
}
\label{fig-separation-TDC-specs}
\end{figure}

Fig. \ref{fig-separation-TDC-specs} shows the Cerenkov separation (as $f_\text{Ceren}$) of 96 photosensors vs. different TDC specifications using the optimal cut value of each photosensor.
The results of 50~ps LSB are not shown in this figure because the min $f_\text{Ceren}$ is 0 regardless of the jitter, meaning no timing cut value allows an efficient separation at some photosensors.
For LSB~$\le$~20~ps, results are nearly identical, the min, median, and max values are approximately 57\%, 66\%, and 75\%, respectively, 
for 96 photosensors.

These results indicate that the time-based separation absolutely requires TDCs with better than 50~ps for the LSB, otherwise the technique will not yield any usable result.
For LSB~$\le$~20~ps, a jitter of 50~ps FWHM is still enough for the time-based separation when using the optimal cut value for each photosensor.
By meeting cut conditions (\ref{eq_cond}), the min $f_\text{Ceren}$ of all photosensors in the time-based separation is still better than the filter-based approach (50\%).
The results also indicate an upper limit when going for smaller jitter and LSB, implying a similarity to raw photon data from LoLX simulation.
Regarding PDC design, these results suggest minor improvements on the separation when LSB~$\le$~20~ps. 

The Cerenkov fraction, $f_\text{Ceren}$, used in Fig. \ref{fig-separation-TDC-specs} is calculated at the optimal cut value of each sensor, using the 5000-event photon time profile.
When calculating for a single event, $f_\text{Ceren}$ greatly fluctuates between 0 and 100\% as there are only few Cerenkov photons per event per sensor.
When considering the arithmetic mean of $f_\text{Ceren}$/event over 5000 events, this mean approaches the value calculated from the 5000-event time profile of each sensor, with a difference of approximately 2\%, within the statistical error of the mean.  
Thus, the $f_\text{Ceren}$ values calculated from the 5000-event time profiles are used for later parts of this paper.

We evaluated the optimal cut values from 6 parameter combinations using jitters (FWHM) of 10 and 20~ps \& LSBs of 5, 10, and 20~ps. 
Of all 6 combinations, the min, median, and max of the optimal cut values are approximately 190~ps, 230~ps, and 305~ps, respectively.
The inter-quartile range is approximately 22.5~ps, around 20\% the range (max~-~min) of 96 photosensors. 
This indicates the optimal cut values for 96 photosensors are not much different from each other, which is expected as the $\beta$ source is located at the center of the LoLX detector cage.


\begin{figure}[!t]
\centering
\includegraphics[width=\linewidth]{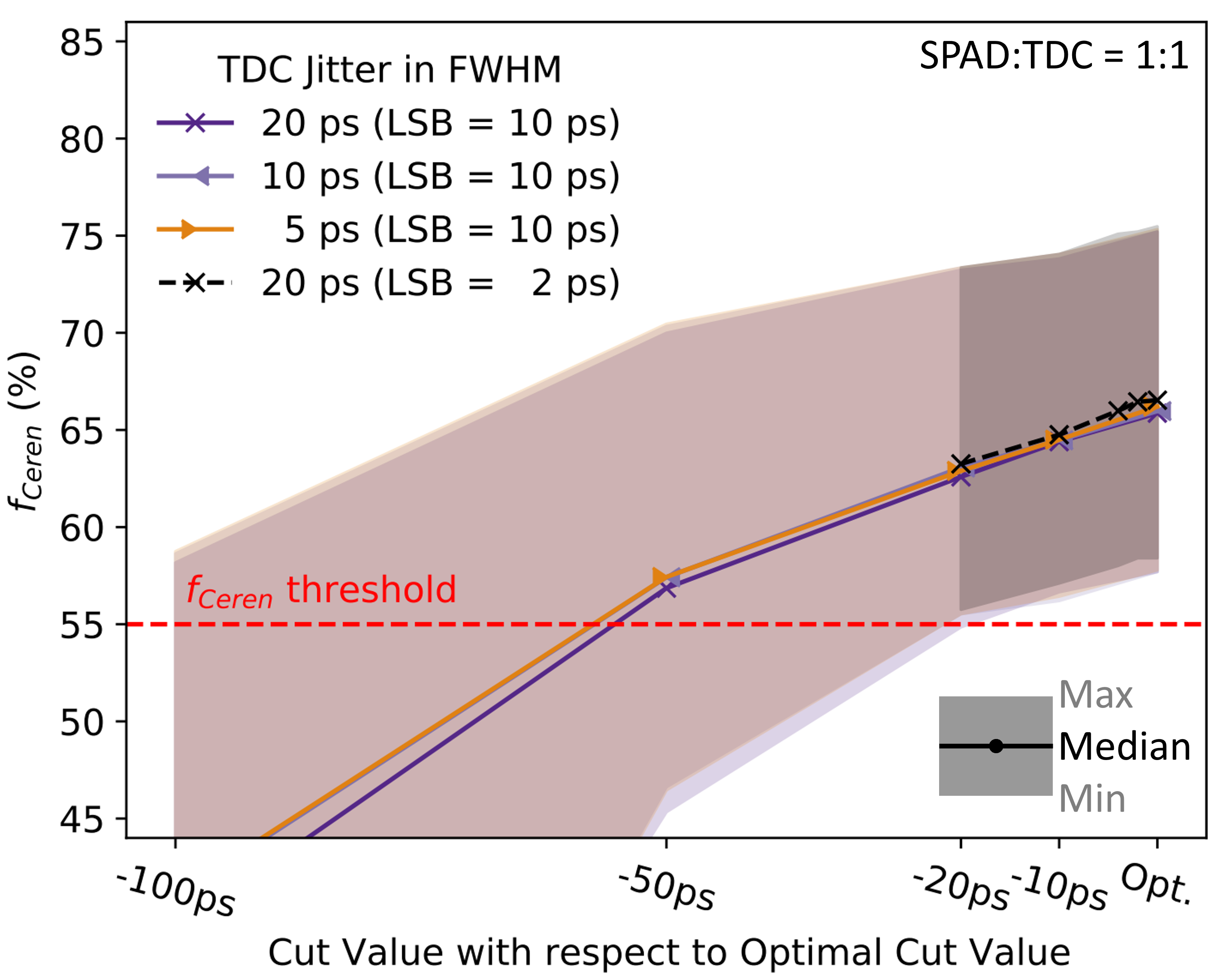}
\caption{Cerenkov separation (as $f_\text{Ceren}$) vs. different cut values with respect to the optimal cut value for each photosensor.
Dots and lines depict medians while shaded areas show min and max values of 96 photosensors.
For the case of 20~ps (FWHM) jitter with 10~ps LSB, with a cut at 20~ps before the optimal value, the min $f_\text{Ceren}$ is slightly below the threshold.
}
\label{fig-separation-cut-pos}
\end{figure}

In addition to the optimal cut value used in Fig. \ref{fig-separation-TDC-specs}, the impact of different cut values on the separation quality was also evaluated.
Fig. \ref{fig-separation-cut-pos} shows Cerenkov separation (as $f_\text{Ceren}$) vs. different cut values with respect to the optimal cut value at four selected configurations.
Within a 50~ps margin before the optimal cut value, at the same cut value relative to the optimal one, reducing jitter (FWHM) from 20~ps to 5~ps has minimal impact on separation performance. 
Moving the cut from the optimal value to 20~ps earlier, the median $f_\text{Ceren}$ of 96 photosensors drops around 3\%.
Earlier than the 50~ps margin, the $f_\text{Ceren}$ drops faster to below the 55\% threshold. 
In all four configurations, with a cut at 20~ps before the optimal value, only the configuration of 20~ps (FWHM) jitter \& 10~ps LSB has the min $f_\text{Ceren}$ slightly below the 55\% threshold due to one photosensor.
Thus, for all 96 photosensors to have at least 3 cut values meeting cut conditions (\ref{eq_cond}), 
it is better to have jitter~$\le$~10~ps FWHM \& LSB~$\le$~10~ps.

Therefore, to ensure time-based separation performance: 1)~LSB~$\le$~10~ps for a sufficient number of cut values, 2)~cut values within a 20~ps margin before the optimal value for an average of 63\% Cerenkov separation or more, and 3)~jitter~$\le$~10~ps FWHM for efficient separation.

\subsection{SPAD:TDC Sharing Ratio} \label{SPAD-TDC-Ratio}
\begin{figure}[!t]
\centering
\includegraphics[width=\linewidth]{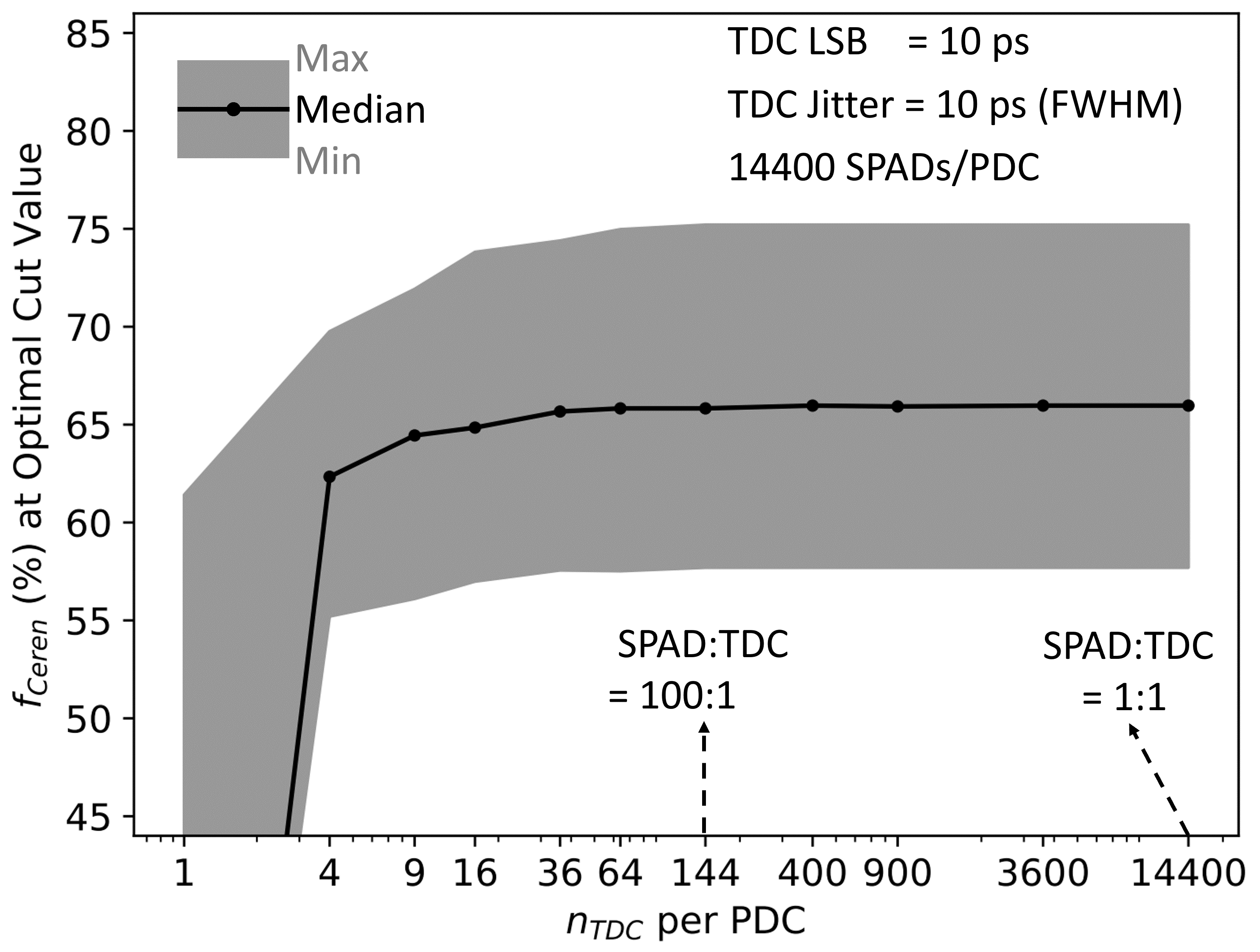}
\caption{
Cerenkov separation (as $f_\text{Ceren}$) of 96 photosensors vs. $n_\text{TDC}$ per PDC with the optimal cut value for each photosensor. 
Each SPAD-TDC group records the timing of the first arrival hit per event.
Dots and lines depict medians while the shaded area shows min and max values of 96 photosensors.}
\label{fig-separation-nTDC}
\end{figure}

To evaluate how the SPAD:TDC sharing ratio affects the separation quality, we used the 10~ps (FWHM) jitter \& 10~ps LSB configuration.
The SPAD array has 120~$\times$~120 cells of \qtyproduct{50x50}{\um}, thus 14400 SPADs per PDC.
Each group of n~$\times$~n SPADs is considered to share a TDC, with n ranging from 1 to 120.
Each SPAD-TDC group records the timing of only the first hit arriving per event.

Fig. \ref{fig-separation-nTDC} shows the Cerenkov separation (as $f_\text{Ceren}$ detected at optimal cut values) vs. the number of TDCs ($n_\text{TDC}$).
From this figure, with 16~TDCs~/ 14400~SPADs (i.e., SPAD:TDC =~900:1), the median $f_\text{Ceren}$ is still 65\%.
The reason is that Cerenkov hits are rare and usually the first hit in each event.
Scintillation hits, mainly arriving later, are less likely than Cerenkov hits to get the first timestamps in same SPAD:TDC group.
Thus, it is possible to separate Cerenkov and scintillation with a modest $n_\text{TDC}$.

\begin{figure}[!t]
\centering
\includegraphics[width=\linewidth]{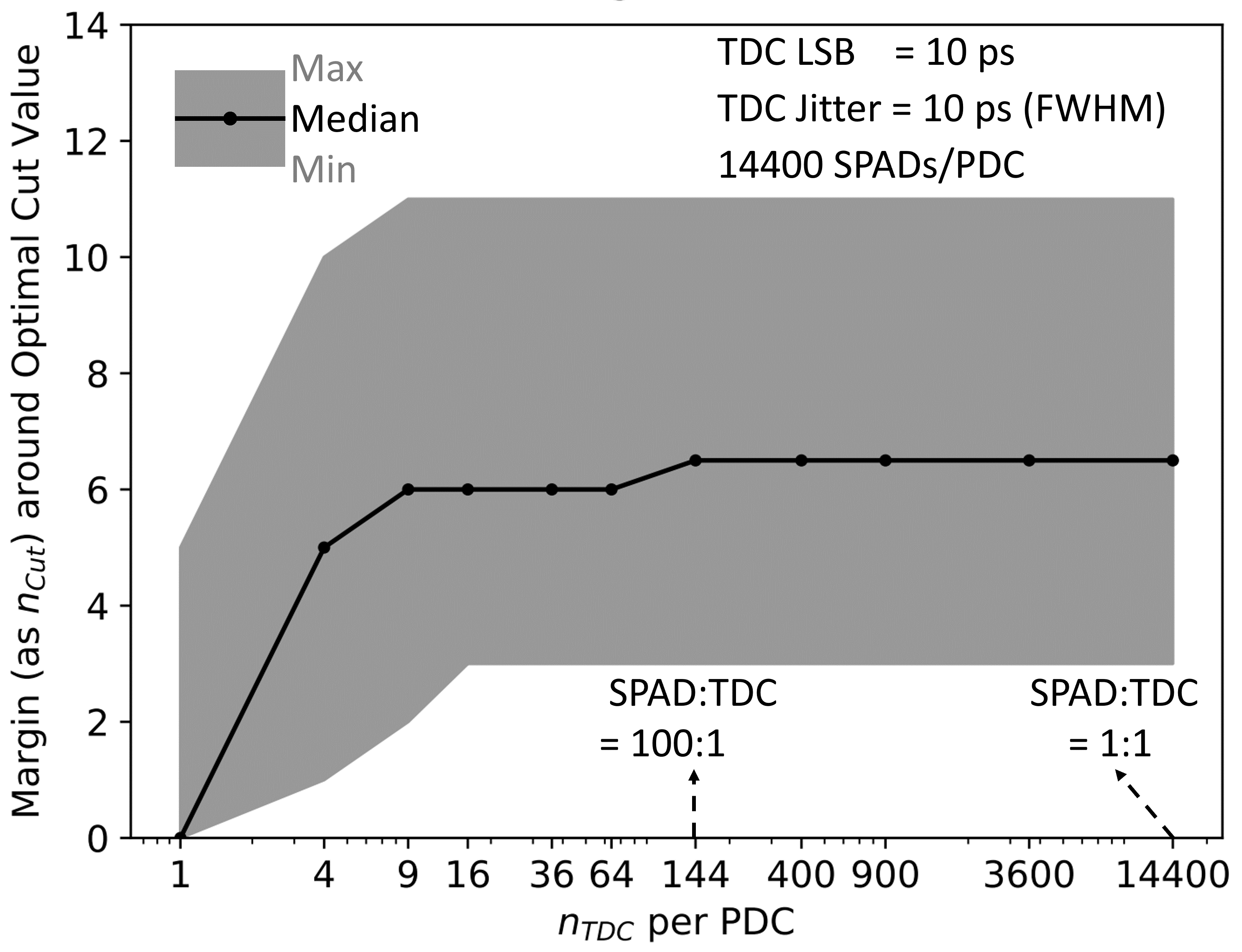}
\caption{
Margin of cut values ($n_\text{Cut}$) meeting conditions (\ref{eq_cond}) around the optimal value vs. $n_\text{TDC}$ per PDC.
The margin is before the optimal value (smaller timestamp direction).
Dots and lines depict medians while shaded areas show min and max values of 96 photosensors.
}
\label{fig_nCut_nTDC}
\end{figure}

Fig. \ref{fig_nCut_nTDC} shows the margin of cut values meeting conditions (\ref{eq_cond}) around the optimal value vs. the $n_\text{TDC}$.
For all photosensors to have a margin of at least 3 cut values meeting the cut conditions (\ref{eq_cond}), i.e. a margin of 20~ps, a minimum of 16~TDCs is needed for 14400~SPADs. 
The exact $n_\text{TDC}$ should be decided from each experiment's requirements, e.g., more TDCs to have more scintillation timestamps for photon profiling or for event arrival time estimators \cite{gundacker2015comparison}, or fewer TDCs to save on power consumption and/or silicon real estate within the sensor.

\subsection{General Discussions}
Photon time profiles of Cerenkov and scintillation are important for the time-based separation modeling. 
As the total $n_\text{Scint}$ is substantially higher than the total $n_\text{Ceren}$, a slightly higher level of fluctuation in the early time window of scintillation can worsen the time-based separation performance.   
A longer decay time of Cerenkov photons can also lead to a similar result. 
This motivates us for more precise timing measurements when PDCs are ready for LoLX in the future.
Experimental results will be reincorporated into the models. 

The results presented in this work are for the $\beta$ source at the center of the LoLX detector cage. 
For different source positions, the optimal cut value of each PDC is expected to shift, reflecting how close the photosensor is to the interaction point.
The performance of the time-based Cerenkov-scintillation separation may not degrade significantly as long as the prompt arrival of Cerenkov photons is maintained.
For better resilience, the overall data should be included to first estimate the interaction position within the instrument volume, after which the optimal cut values for each photosensor may be adjusted, and will be the topic of future studies. 

\section{Summary \& Conclusions} \label{conclusions}
In this work, a PDC simulator, whose input is the light information from the LoLX simulation model, was used to study the effect of TDC parameters on the time-based Cerenkov separation performance.
This study aims to open specification margins for the ongoing development of LoLX PDCs, which, in the future, will replace current SiPMs used in LoLX.

TDC jitter and LSB results show that the time-based separation improves the Cerenkov detection on average from 50\% (optical filters) to 66\%. 
This 66\% is influenced by the Cerenkov and scintillation photon time profiles, 
whose fine characterization is one of LoLX's main research goals. 

A configuration of 10~ps (FWHM) jitter \& 10~ps LSB is expected to give a cut margin of at least 3 values (20~ps before the optimal) for all 96 photosensors with an average Cerenkov detection ranging from 63\% (at 20~ps before optimal) to 66\% (at optimal).
The SPAD:TDC sharing ratio results show that around 65\% the total $n_\text{Ceren}$ can be detected with a modest 16 TDCs per PDC.
%

The above findings offer a specification outline, specifically on TDC jitter, LSB, and SPAD:TDC sharing ratio, to define LoLX PDC development.
Once these PDCs are ready, LoLX will use them to experimentally verify these simulations.
In the long run, the LoLX results will provide valuable insights for large-scale LXe-based experiments \cite{adhikari2021nexo, BRODSKY201976} and pave the way for the future development of a LXe-based TOF-PET demonstrator system \cite{PETALO2020, doke-2006-TOF-PET-LXe}.

\section*{Acknowledgments}
The authors thank Julien Roy-Sabourin and David Paré for their help on matching LoLX simulation output with DSAS input.
The authors also thank Gabriel Bélanger, Julien Rossignol, Xavier Groleau, and Audrey Corbeil Therrien for their help on using the DSAS package.

\bibliographystyle{IEEEtran}
\bibliography{NSS-MIC-2022}

\end{document}